\documentclass[aip,apl,reprint,superscriptaddress]{revtex4-1}

\usepackage{soul}
\usepackage{ulem}

\usepackage{graphicx}
\usepackage[ansinew]{inputenc}
\usepackage{array}
\usepackage{color}
\usepackage{amsmath}
\usepackage{amsxtra}
\usepackage{amstext}
\usepackage{amssymb}
\usepackage{latexsym}
\begin{document}

\title{Optical thickness determination of hexagonal Boron Nitride flakes}

\author{Dheeraj Golla}
\affiliation{Department of Physics, University of Arizona, Tucson, AZ, 85721 USA.}
\author{Kanokporn Chattrakun}
\affiliation{Department of Physics, University of Arizona, Tucson, AZ, 85721 USA.}
\author{Kenji Watanabe}
\affiliation{National Institute for Materials Science,1-1 Namiki, Tsukuba,305-0044 Japan.}
\author{Takashi Taniguchi}
\affiliation{National Institute for Materials Science,1-1 Namiki, Tsukuba,305-0044 Japan.}
\author{Brian J. LeRoy}
\affiliation{Department of Physics, University of Arizona, Tucson, AZ, 85721 USA.}
\author{Arvinder Sandhu}
\affiliation{Department of Physics, University of Arizona, Tucson, AZ, 85721 USA.}
\affiliation{College of Optical Sciences, University of Arizona, Tucson, AZ, 85721 USA.}

\date{\today}

\begin{abstract}
Optical reflectivity contrast provides a simple, fast and noninvasive method for characterization of few monolayer samples of two-dimensional materials. Here we apply this technique to measure the thickness of thin flakes of hexagonal Boron Nitride (hBN), which is a material of increasing interest in nanodevice fabrication. The optical contrast shows a strong negative peak at short wavelengths and zero contrast at a thickness dependent wavelength. The optical contrast varies linearly for 1-80 layers of hBN, which permits easy calibration of thickness. We demonstrate the applicability of this quick characterization method by comparing atomic force microscopy and optical contrast results.
\end{abstract}

\maketitle
Hexagonal Boron Nitride (hBN) has a planar hexagonal structure similar to graphite and has proven to be an excellent substrate for graphene based electronic and opto-electronic devices. It has been shown that graphene devices on hBN substrates show enhanced performance like increased carrier mobility and reduced charge fluctuations\cite{Dean:2010jy,Xue:2011dv,Decker:2011wh}. Graphene sheets conform to atomically flat hBN resulting in reduced roughness and charge puddle formation as compared to other common substrates such as Si/SiO$_{2}$\cite{Xue:2011dv,Decker:2011wh}. Thin hBN flakes have also proven to be an excellent dielectric or tunnel barrier for device applications\cite{Taychatanapat:2011hr,Ponomarenko:2011cj,Britnell:2012jp,Gorbachev:2012bn} and can modify graphene's band structure \cite{Yankowitz:2012gi}. A large direct bandgap makes hBN attractive for compact UV laser applications\cite{Watanabe:2004ct}. Interestingly, the success of hBN in graphene electronics is now also being mirrored in the development of other two dimensional materials such as transition metal dichalcogenides (TMD) (e.g. MoS$_2$, MoSe$_2$, WS$_2$ etc.) devices, where hBN substrates have led to 10 times better photoluminescence quantum yields than Si/SiO$_2$.\cite{Mak:2012tf}

It is anticipated that hBN will be an essential constituent for future graphene and TMD heterostructure devices in many roles ranging from a tunnel barrier to a gate dielectric. Hence, it is very important to have methods for quick, economical and non-invasive characterization of hBN flakes, specifically, the exact number of hBN monolayers and its flatness over the size of the device. In that regards optical reflection microscopy has proven to be a highly useful tool. Optical contrast measurement have been used to identify mono and few layered graphene on various substrates\cite{Blake:2007hb,Gaskell:2009bh,Wang:2009cu}. Identification of monolayer and bilayer hBN has also been reported using reflectivity contrast\cite{Gorbachev:2010vg}. In this paper, we characterize hBN flakes deposited on a SiO$_{2}$/Si substrate. We establish parameters which can enable quick identification for hBN flakes varying from a few to 100 layers. We also show that this approach is sensitive to optical thickness changes as small as 1-2 layers allowing the identification of steps in flakes which appear flat under white light illumination. 

\begin{figure}[ht]
\includegraphics[width=0.43 \textwidth]{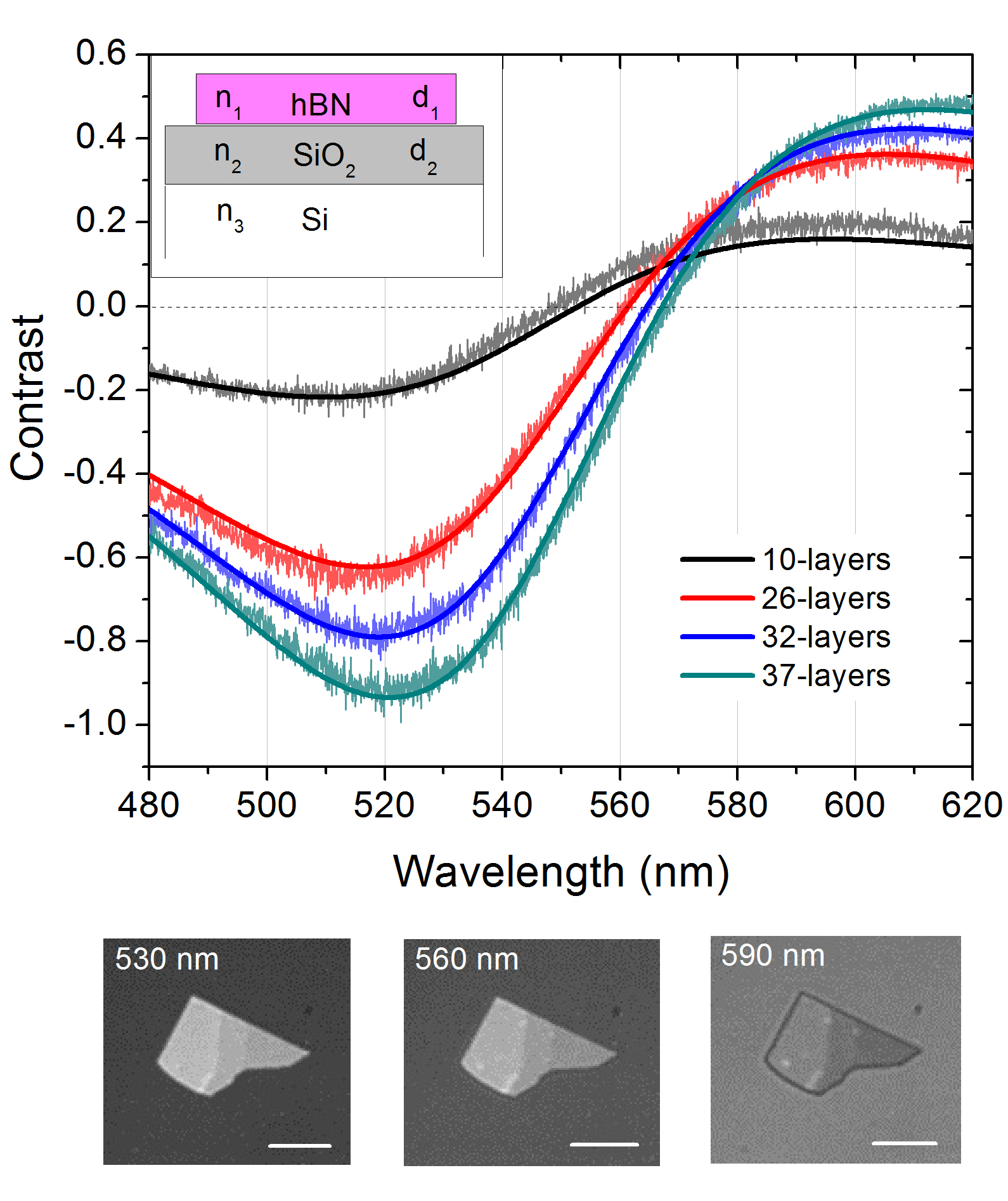}
\caption{\label{fig:1} The optical contrast of hBN on SiO$_2$/Si substrate as a function of the wavelength of light. Different curves corresponds to flakes with different number of layers. The dark solid lines are the fits obtained using the multilayer interference model discussed in the text. The bottom panels show optical microscope images at three discrete wavelengths corresponding to positive, zero and negative contrast.  The scale bar represents 5 $\mu$m.}
\end{figure}

The sample geometry is shown schematically in the inset of figure \ref{fig:1}.  Few layer hBN flakes were prepared by mechanical exfoliation of hBN single crystals on 285 nm SiO$_2$/Si substrates.  The hBN was grown using the high-pressure method that has been previously described\cite{Taniguchi:2007hl}. Using a femtosecond Ti:Sa amplifier and a filamentation setup\cite{Roberts:2009kr}, we nonlinearly broaden the light pulse spectrum to produce a broadband white light source spanning 460-850 nm. White light is incident normally from air on the hBN-SiO$_2$-Si structure and is focused to a small spot size using a 50x, NA=0.5 microscope objective. The backwards reflected light is collected and after passing through a beamsplitter it is imaged on to the entrance slit of a spectrometer. The spectrally resolved reflection signal is obtained using a CCD at the exit aperture of the spectrometer. 

The quantity measured in our experiments is the optical contrast in reflectivity of a three tiered structure, which can be defined as 
\begin{equation}
C =\frac{R_{SiO_{2}}-R_{hBN+SiO_{2}}}{R_{SiO_{2}}}
\end{equation}
where $R_{SiO_{2}}$ is the reflection coefficient at normal incidence for a bare SiO$_2$/Si substrate and $R_{hBN+SiO_{2}}$ is the same for an hBN covered substrate.  This corresponds to the normalized change in reflectivity of the hBN compared to the substrate.

The measured optical contrast is plotted in figure \ref{fig:1} as a function of wavelength for hBN flakes of varying thicknesses. The thickness of the hBN flakes was determined after optical imaging using atomic force microscopy.  A thickness of 3.33 nm corresponds to 10 layers of hBN because the c-axis lattice constant of hBN is 0.333 nm. Several observations can be made from this data. The contrast for a given flake can be both positive and negative depending on the wavelength, with a zero crossing in between. A contrast of zero would mean the flake being invisible at that wavelength, i.e. it has the same reflectivity as the substrate. The zero contrast wavelength is dependent on the flake thickness and varies from 550 nm to 570 nm (for 282 nm SiO$_2$) as the number of layers increases from a few to tens of layers. The negative contrast is stronger and its peak is observed between 520-530 nm. The positive contrast peaks between 590-620 nm. On the bottom panels of figure \ref{fig:1} we show three microscope images of a hBN flake taken at selected wavelength bands centered on 530, 560 and 590 nm. Clearly, one can see the flake has greater reflectivity than substrate at 530 nm, i.e. negative contrast, and lesser reflectivity than substrate, i.e. positive contrast at 590 nm. The 560 nm image corresponds to the region near the zero contrast wavelength. Since, the flake has two step edges, there are three distinct thicknesses that can be observed. The sharp tip of the flake is thinnest and hence closest to  zero contrast at 560 nm.

\begin{figure}[ht]
\includegraphics[width=0.45 \textwidth]{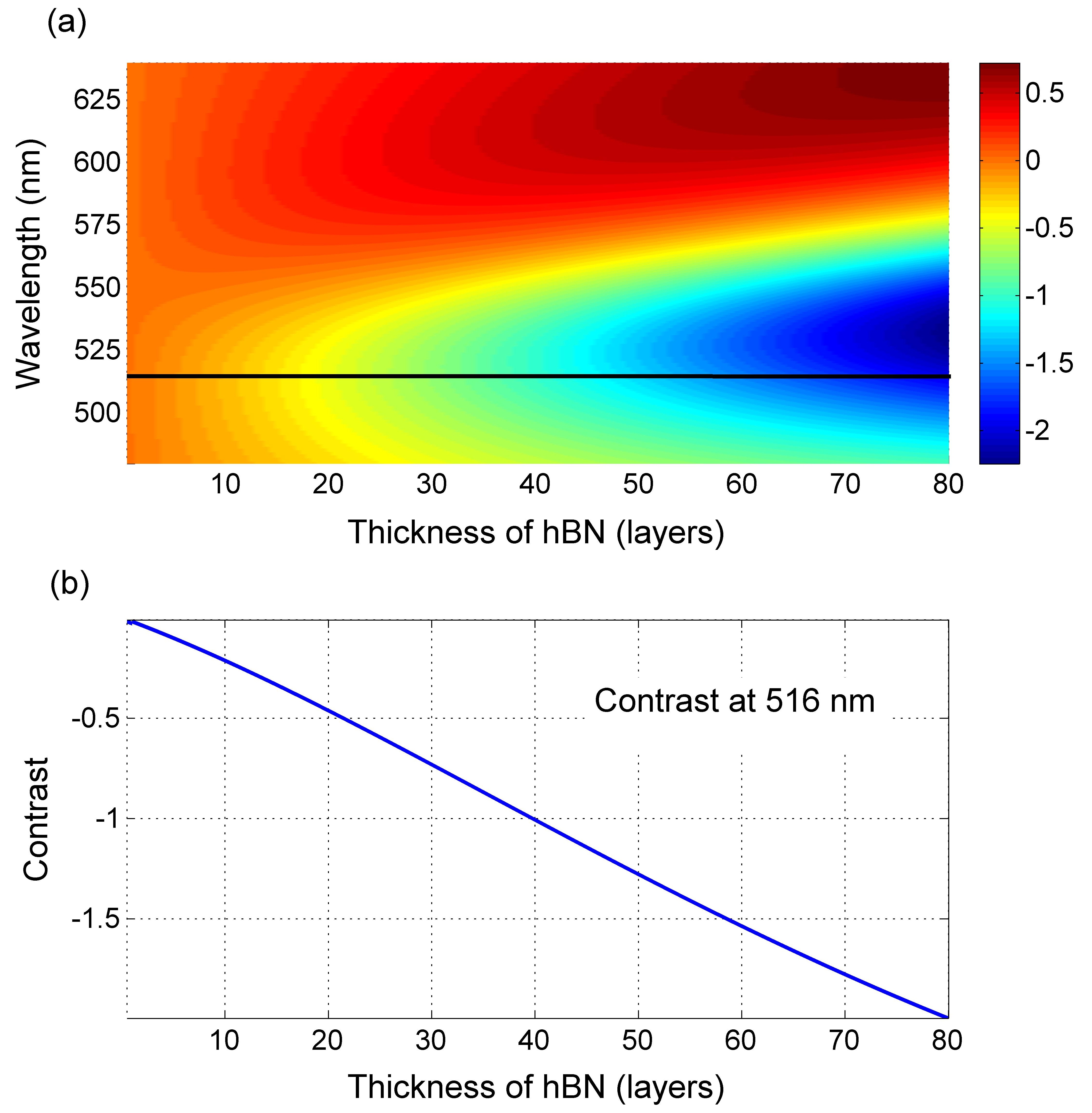}
\caption{\label{fig:2}(a) Calculated optical contrast a function of the wavelength of light and thickness of hBN for a 282 nm SiO$_2$ layer. (b) Line out at 516 nm shows that the contrast varies  linearly with the thickness of the hBN flake between 1 and 80 layers.}
\end{figure}

In order to model the optical contrast we use a standard multilayer interference approach\cite{Hecht}. Essentially, a  transfer matrix method is used, where we consider three layers hBN, SiO$_2$, and Si having refractive indices $n_1$, $n_2$, and $n_3$ respectively (refer to the inset of figure \ref{fig:1}) . The light is incident from the airside which implies $n_0$=1. A recursive relation can be written in general for the electric fields at the interfaces as  
\begin{equation}
\Gamma_j = \frac{r_j+\Gamma_{j+1}exp(-2ik_{j}d_{j})}{1+r_j\Gamma_{j+1}exp(-2ik_{j}d_{j})}
\end{equation}
where $d_j$ is the thickness of layer $j$, $\Gamma_j = E_{j-}/E_{j+}$ with $E_{j-}$ and $E_{j+}$ representing the electric field on either side of $j^{th}$ interface, and 
$$r_{j}=\frac{n_{j-1}-n_{j}}{n_{j-1}+n_{j}}$$
is the reflection coefficient of the $j^{th}$ interface. The wavevector  $k_{j}=2\pi n_{j}/\lambda$ and the thickness of the hBN and SiO$_2$ are $d_1$ and $d_2$, while the thickness of Si is assumed infinite. The recursion is initiated at the SiO$_{2}$/Si interface by setting $\Gamma_3=r_3$ and then $\Gamma_2$ and $\Gamma_1$ are successively calculated to obtain reflectivity of the stack $R = |\Gamma_1|^2$. One can easily generalize this method to a stack with an arbitrary number of interfaces.

The contrast between the region with a hBN flake and the background SiO$_2$ can then be evaluated by setting the refractive index $n_1$ of hBN layer equal to 1 (air) and using $C = (R(n_1=1)-R(n_1=hBN))/R(n_1=1)$. The refractive index of hBN is taken to be 1.85 at 560 nm, and a weak linear variation is invoked, leading to a 3\% higher (lower) value at 480 (640) nm \cite{Stenzel:1996}. Both SiO$_2$ and the underlying Si have wavelength dependent complex refractive indices\cite{Palik}.

The solids curves in figure \ref{fig:1} show the calculated optical contrast for various hBN thicknesses. The thickness of the SiO$_{2}$ layer was taken to be 282 nm in all of the calculations, which is within the $285\pm5$ nm value listed by the manufacturer for the thickness of the SiO$_2$ on the Si wafer.  The thickness of the hBN flakes was determined by AFM measurements.  The fits to the experimental curves are very good over the entire wavelength range.

In figure \ref{fig:2} (a) we show the dependence of the optical contrast on the wavelength of light as well as the thickness of the hBN layer for 1 to 60 layers. This calculation is performed with a 282 nm thick SiO$_2$ layer to match the experimental data.  The negative contrast peak increases rapidly in intensity with the number of layers. In figure \ref{fig:2} (b) we show the variation of contrast at a fixed wavelength of 516 nm which is near the negative peak. It can be seen that the contrast varies linearly with the thickness of hBN from 1 to 80 layers. This fact implies that contrast observation at a single wavelength can be used for fast and accurate characterization of the thickness of the hBN flake.  In particular, we find that the contrast varies by $\sim$2.5\% per layer of hBN at this wavelength.

\begin{figure}[h]
\includegraphics[width=0.49 \textwidth]{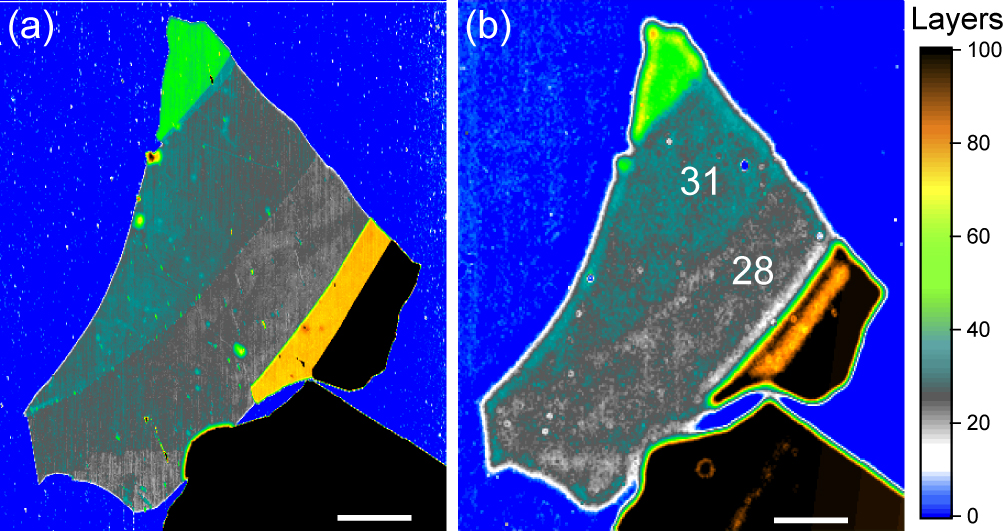}
\caption{\label{fig:3} The comparison of optical contrast image with the atomic force microscopy image. (a) AFM image of a hBN flake on an SiO$_2$ substrate.  (b) Optical contrast image of the same hBN flake.  The thickness of the two central regions in number of layers is marked on the image.  Both images use the same color scale.  The optical contrast values are converted to thickness using 2.5\% per layer.  The scale bar represents 2 $\mu$m. The comparison illustrates the ability of the optical contrast method to measure the thickness of hBN flakes as well as differentiate small steps corresponding to changes in thickness of a few layers of hBN.}
\end{figure}

We show a practical application of this imaging method in figure \ref{fig:3}. Figure \ref{fig:3}(a) shows an atomic force microscope (AFM) image of an hBN flake which exhibits many different thicknesses ranging from 28 to 81 layers. In particular, there are two distinct regions in the center which are separated by a small step of 3 layers.   The typical acquisition time for an AFM image with this resolution is on the order of 10 minutes.  In figure \ref{fig:3}(b) we show an optical contrast image of the same flake obtained at 516 nm wavelength using an optical microscope. A bandpass filter centered at 516 nm and with a width of 5 nm was placed in front of the microscope light in order to acquire the image.  The resulting image was recorded on a monochrome CCD camera with an exposure time of 0.25 s.  One can clearly observe the 3 monolayer step in the optical contrast image with over a factor of 1000 improvement in the time required for imaging the flake. 

In conclusion, we have used an optical method to characterize the thickness of hBN flakes ranging from a few layers to tens of layers. We find that the wavelength dependence of the contrast exhibits three unique regions which can be used to calibrate the thickness directly. For a SiO$_2$ substrate thickness of 282 nm, these regions are 515-530 nm for large negative contrast, 550-570 nm for near zero contrast, and 590-620 nm for positive contrast. The negative contrast varies linearly with the thickness of hBN from 1 to 80 layers. This provides a straightforward thickness calibration using only optical methods. We verify this calibration method using an AFM image of a hBN flake which shows variable thickness including a small step of three layers. The ability to easily characterize the hBN flake thickness as well as differentiate small changes in thickness is expected to be extremely helpful in assembly of graphene and TMD devices based on the use of hBN as a substrate or an dielectric insulating layer.

K.C. and B.J.L. acknowledge support from the U. S. Army Research Office under contract W911NF-09-1-0333.

\bibliography{Golla_HBN_final}


\end{document}